\documentclass{aanew} 
\usepackage{graphicx} % 3/2001
\usepackage{natbib} % 3/2001
\bibpunct{(}{)}{;}{a}{}{,} % 3/2001, aadoc.pdf
\begin{document}
\title { Optimization of radio astronomical observations using Allan
  variance measurements } \author {R. Schieder, C. Kramer}
\offprints{R. Schieder} \institute{ I. Physikalisches Institut,
  Universit\"at zu K\"oln,
  Z\"ulpicher Stra\ss e 77, 50937 K\"oln, Germany\\
  \email{schieder@ph1.uni-koeln.de} } \date {Received /
  accepted 01.05.2001} \abstract{ Stability tests based on the Allan
  variance method have become a standard procedure for the evaluation
  of the quality of radio-astronomical instrumentation. They are very
  simple and simulate the situation when detecting weak signals buried
  in large noise fluctuations. For the special conditions during
  observations an outline of the basic properties of the Allan
  variance is given, and some guidelines how to interpret the results
  of the measurements are presented. Based on a rather simple
  mathematical treatment clear rules for observations in
  ``Position-Switch'', ``Beam-'' or ``Frequency-Switch'',
  ``On-The-Fly-'' and ``Raster-Mapping'' mode are derived. Also, a
  simple ``rule of the thumb'' for an estimate of the optimum timing
  for the observations is found. The analysis leads to a conclusive
  strategy how to plan radio-astronomical observations. Particularly
  for air- and space-borne observatories it is very important to
  determine, how the extremely precious observing time can be used
  with maximum efficiency. The analysis should help to increase the
  scientific yield in such cases significantly.
} % abstract
\maketitle

%\thesaurus{(03.09.3; % Instrumentation: miscellaneous
%            03.13.2; % Methods: data analysis
%            03.13.5; % Methods: observational
%            03.19.3; % Space vehicles: instruments
%            03.20.8; % Techniques: spectroscopic
%            03.20.9)} % Telescopes

%% till here ok with new style file
%% --------------------------------

%%% hier klappt es mit Figure. 1

\section{Introduction}

Allan variance measurements have been demonstrated as a useful tool
for the characterization of the stability of radio-astronomical
equipment such as Millimeter or Submillimeter-receivers or large
bandwidth back-ends \citep{schieder1985,kooi2000}. Particularly for the
development of acousto-optical spectrometers (AOS) at the K\"olner
Observatorium f\"ur Sub-Millimeter Astronomy (KOSMA) the method has
played a very important role, because it provides clear evidence that
the spectrometers are well suited for the use at an observatory by
means of a reliable test laboratory procedure \citep{tolls1989}. The simple
definition of the Allan variance makes it very easy to apply such
measurements also for the characterization of the stability of other
instruments, a very elementary case is the definition of the quality
of a simple Lock-In amplifier for example.

For a real time spectrometer, as used in radio-astronomy with many
simultaneously operating frequency channels, it is a very important
condition that all channels are behaving identically in a statistical
sense. Therefore, the use of the Allan variance for the investigation
of the performance of the spectrometer is based on the assumption that
there are no differences between different frequency channels. That
this is not always correct is evident. Thus, it is always necessary to
verify the similarity of all frequency channels of the spectrometer by
investigating the baseline noise of measured spectra for example.
Typical problem areas for instance are light scatter problems in
acousto-optical spectrometers (AOS), where speckles may affect
individual channels more heavily than others. The same is true for
filterbanks which have occasionally same peculiar channels even in a
well maintained back-end system. But in all normal cases of well
behaved instrumentation, the Allan variance plot is a most useful
method to precisely characterize the instrumentation in use.

In general, observations at an observatory are done with the available
instrumentation as is, and it can not be modified or even improved by
the observer. On the contrary, the observer has to find the correct
observing parameters in order to use the available hardware in a most
economic way. It is the purpose of this paper to develop a strategy
for an optimization of the observing process. For this the knowledge
of the stability parameters is decisive. Once this information is
available from an Allan variance measurement for example, it should be
a rather straightforward matter to determine the essential parameters
like length of integration per position on sky et cetera. The
following mathematical treatment analyses the commonly used observing
methods, i.e. ``Po\-sition-'', ``Beam-'' or ``Frequency-Switch'',
``On-The-Fly'' (OTF) measurements or ``Raster-Mapping'' based on the
information contained in the Allan variance plot. As a result
practical guidelines for the most efficient observing method are
found, which can be used at any radio observatory. Particularly, all
space- or air-borne observatories require a most efficient use of the
extremely precious observing time, since any loss can usually not be
compensated by a simple increase in observatory time. But also for
ground-based observatories the results found in the following should
be very useful.

\section{Definition of the Allan variance}

If a test procedure is defined for use at any time and at any
location, it needs to be as simple and unique as possible. Therefore,
we understand the Allan variance as the ordinary statistical variance
of the difference of two contiguous measurements (see also
\citet{rau1984}). One has to consider a signal-function s(t), which is
the instantaneous output signal of a spectrometer channel or of a
continuum detector for example. The output is now integrated for a
time interval T representing an estimate of the mean signal which is
stored as spectrometer data in the computer:

\begin{equation}
  x(T,t) = 1/T \int_{t-T}^t s(t') dt' 
\end{equation}

The expectation value of $x(T,t)$ is therefore identical with the
expectation of $s(t)$. For the observation of weak signals, a certain
number $N$ of differences of two of these data, a
``signal-measurement'' $x_{s}$ and a ``reference-measurement''
$x_{r}$, are subtracted from each other:
\begin{equation}
  d = x_{s} - x_{r}
\end{equation}
so that the desired signal alone becomes visible when
averaging. Typically, each of the two measurements are done at
different times, after the telescope has moved between two positions
on sky.

In order to obtain a plausible estimate of the error of the difference
we use the standard definition of the variance:

\[
  \sigma_d^2(T) = \langle (d-\langle d\rangle)^2 \rangle
                = \langle d^2 \rangle - \langle d \rangle^2
\]
The brackets "$\langle \rangle$" stand for the expectation value. In
comparison, this definition is similar to the original definition of
the Allan variance \citep{allan1966}, if one considers a situation,
where the expectation value of the difference is zero which is
practically ``normal'' during radio-astronomical observations: 

\[
  \sigma_{\rm A}^2(T) = 1/2 \langle d^2 \rangle.
\]

For further treatment we use the standard definition of the variance,
but leave the factor of $1/2$ in place for historical reasons, since
it was already introduced by Allan in 1966. Thus we use\footnote{ This
original definition through the difference of samples may be altered
by using the ratio of contiguous data instead. The corresponding
"ratio-variance" is then: $\sigma_{r}^2(T) = 1/2 \times \langle
[x_{s}/x_{r}-\langle x_{s}/x_{r}\rangle]^2 \rangle$.

In case the rms of the noise is small as compared with the
mean $\langle s(t)\rangle$, one can easily show that
$\sigma_{r}^2(T) = \sigma_{\rm A}^2(T)/\langle s(t) \rangle^2$.
This new definition has the advantage to properly calibrate the data
even at varying gain in the system. }:

\begin{equation}
  \sigma_{\rm A}^2(T) = 1/2 \langle (d-\langle d\rangle)^2 \rangle
                = 1/2 [\langle d^2 \rangle - \langle d \rangle^2]
\end{equation}

Note that with this new definition we consider also the possibility
that the mean of the difference may not be zero. In case there is
radiometric noise only, this expression defines the noise of a single
measurement $x_{s}$ or $x_{r}$ alone thanks to the factor of
$1/2$.\footnote{ In general one has to consider the fact that there is
only a finite data set available for the calculation of a
variance. Therefore, instead of Eq.(3), one should use the standard
definition with
$\sigma_{\rm A}^2(T)=\frac{1/2}{N-1} \sum_{n=1}^{N}(d_n-\overline{d})^2$ 
with $\overline{d} = 1/N \sum_{n=1}^{N}d_n$.}

If we apply now Eq.(1), we get:

\[
 \sigma_{\rm A}^2 = [\sigma_{s}^2(T)+\sigma_{r}^2(T)]/2 - 
              [\sigma_{s}^2(T) \sigma_{r}^2(T)]^{1/2} g_{sr}(T)
\]
with
\begin{eqnarray}
 g_{sr}(T) & = & \frac{\langle(x_{s}-\langle x_{s}\rangle)
                   (x_{r}-\langle x_{r} \rangle) \rangle}
 {[\langle(x_{s}-\langle x_{s}\rangle)^2\rangle \langle(x_{r}-\langle
  x_{r}\rangle)^2\rangle]^{1/2}}, \\
 \sigma_{s}^2(T) & = & \langle(x_{s}-\langle x_{s}\rangle)^2\rangle
 \,\,{\rm{and}}\,\,
 \sigma_{r}^2(T) = \langle(x_{r}-\langle x_{r}\rangle)^2\rangle \nonumber. 
\end{eqnarray}

$g_{sr}(T)$ is the normalized cross-correlation function of the two
data sets $x_{s}$ and $x_{r}$. It should be understood that the
expectation values are the means averaged over the time $t$. In other
cases it might be the mean of a large number of spectrometer pixels
for example. Both cases should be equivalent for the discussion here.

If we have the same statistics for both, ``$s$'' and ``$r$'' 
$(\sigma_{r}^2(T)=\sigma_{s}^2(T)=\sigma^2(T))$, then we get finally:

\[
  \sigma_{\rm A}^2(T) = \sigma^2(T) [1 - g_{sr}(T)]
\]

According to this expression the Allan variance is always smaller than
the normal variance of the data sets as long as there is no
``anti-correlation'' with negative $g_{sr}(T)$. The measurement of
differences therefore removes all contributions from the noise which
are correlated. This reflects the simple fact that the impact of slow
drift noise on the signal to noise ratio can be removed by signal
modulation techniques, as is commonly applied during observations in
radio-astronomy or when using Lock-In amplifiers in laboratory
experiments. It also tells immediately that fast switching does not
help whatsoever, if there is no correlation as is typical for pure
white noise.

We have not yet made any particular assumption about the source of the
signal- and the reference-data. For our application here, the two data
``$s$'' and ``$r$'' are derived from the same output signal $s(t)$ of one
spectrometer channel. The two acquisition periods of length T for the
integration of $x_{s}$ and $x_{r}$ must therefore occur one after the other in
order to avoid any undesirable overlap between the two
measurements. For an unequivocal definition of the instrumental Allan
variance we assume that all ``$s$'' and ``$r$'' measurements are contiguous
without any dead time in between. In real life, when observing, there
will be always some unavoidable dead time, since the telescope needs
to be moved between the On- and the Off-position or there is time
needed for data transfer etc. Any delay will increase the impact of
slow drift noise, and it will therefore result in a different
appearance of the system noise. Such effects will be discussed in the
next chapter.

%%% Fig.1-Probe:
\begin{figure*} [htb]
% \resizebox{\hsize}{!}{\includegraphics{test.eps}}
\resizebox{\hsize}{!}{\includegraphics{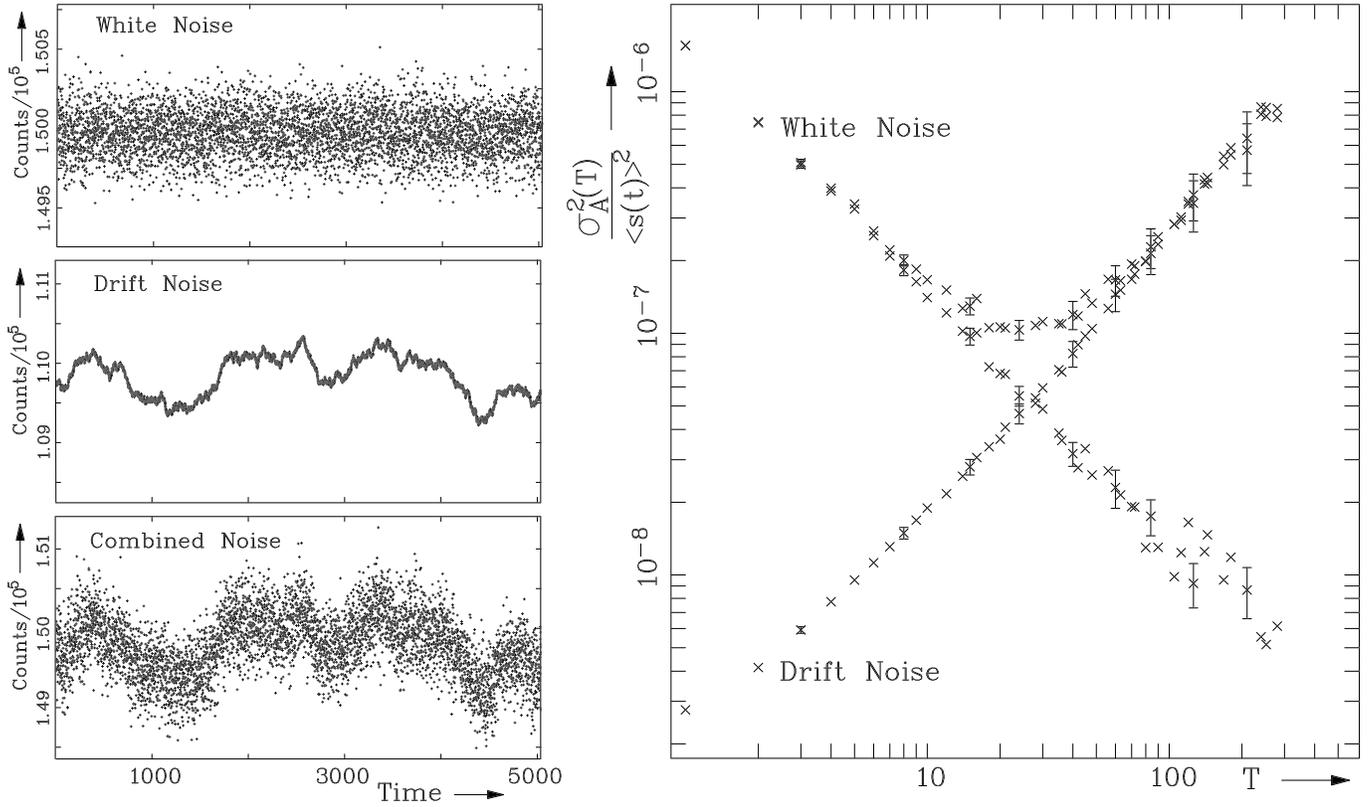}}
% \resizebox{6cm}{!}{\includegraphics{ci_map_spec_x.eps}}
\caption{ \label{FIG_SPECTRA1}
  Artificial data set generated by random numbers (left) with white
  noise of Gaussian distribution (top), drift noise (middle), and
  combined noise (bottom). Each data point corresponds to a sample
  integrated for 1 second while the fluctuation bandwidth was set to
  600 kHz. The drift noise is calculated by filtering white noise with
  a sufficiently broad boxcar time-filter (width $> T_{\rm{max}}$ in
  the Allan variance plot). To the right the (relative) Allan variance
  plots of all three noise spectra are depicted. The white noise
  appears with a slope of $-1$, the drift noise with a slope of
  approximately $+1$. The combination of both results in a typical
  Allan plot with a minimum at some fairly well defined minimum time.
  }
\end{figure*}

\section{The role of the minimum}

For a given integration time the signal output of one spectrometer
channel is described by Eq.(1). We can describe the instantaneous
noise signal $s(t)$ before integrating using the (in this case not
normalized) auto-correlation function $\gamma(\tau)$, but here as a
function of delay time $\tau$:

\begin{equation}
  \gamma(\tau) = \langle (s(t+\tau)-\langle s(t+\tau)\rangle)
  (s(t)-\langle s(t)\rangle)\rangle.
\end{equation}

The integrated signals ``$x$'' have a new auto-correlation function
$\Gamma_T(\tau)$:

\[
  \Gamma_T(\tau) = \langle(x(T,t+\tau)-\langle x(T,t+\tau)\rangle)
                   (x(T,t)-\langle x(T,t)\rangle)
                   \rangle,
\]
and we get after some manipulation, when using Eq.(1) and (5):

\begin{equation}
  \Gamma_T(\tau) = 1/T^2 \int_{-T}^{T}
                   (T-|t|) \gamma(t+\tau) dt.
\end{equation}

According to the definition of the Allan variance in Eq.(3) we have now:

\begin{equation}
  \sigma_{\rm A}^2(T) = \Gamma_T(0) - \Gamma_T(T)
\end{equation}

Frequently, instead of the auto-correlation function $\gamma(\tau)$,
the noise power spectrum $S(f)$ is used for the description of
noise. Since the signal $s(t)$ is real valued, one can write (see
e.g. also in \citet{barnes1971,vessot1976}):
\begin{eqnarray}
  \gamma(\tau) & = & \int_0^\infty S(f) \cos(2\pi f \tau) df 
  \,\,\,{\rm{and}} \\
  S(f) & = & 4 \int_0^\infty \gamma(\tau) 
         \cos(2\pi f \tau) d\tau \nonumber
\end{eqnarray}

How the correlation function behaves in low order approximation for a
noise spectrum like $S(f)\propto1/f^\alpha$ is easily found using
Eq.(8) for sufficiently small $\tau>0$:

\begin{equation}
\begin{array}{rcll}
 \gamma(\tau) & = & g_{c} - g_\alpha \tau^{\alpha-1} 
  & {\rm{for}} \,\,1<\alpha\le3, \\
              & = & g_{c} - g_1 \log(\tau)
  & {\rm{for}} \,\,\alpha=1 \,\,{\rm{(flicker \,\,noise)}} \nonumber\\
              & = & g_{c} + g_\alpha 1/\tau^{1-\alpha}
  & {\rm{for}} \,\,0<\alpha<1 \nonumber\\
              & = & g_0 \delta(\tau)
  & {\rm{for}} \,\,\alpha=0 \,\,{\rm{(white \,\,noise)}}.
\end{array}
\end{equation}

The parameters $g_{c}, g_\alpha, g_1, g_0$ describe the actual
contribution to the correlation function. In all cases we have:
$\gamma(-\tau)=\gamma(\tau)$. According to Eq.(5), $\gamma(0)$ is
identical with the expectation value of the square of the signal,
which is equivalent to the total power contained in the noise
fluctuations, and it has to be finite. Consequently, $1/f^\alpha$
power spectra also have to stay finite at frequencies close to zero,
at least for $\alpha\ge1$, since the integral over the noise power
spectrum $S(f)$ for zero $\tau$ must not diverge for the same reason
(see Eq.(8) for $\tau\rightarrow0$). It means that $1/f^\alpha$
spectra cannot exist at very small $f$! It is easy to deal with the
divergence problem by introducing a lower cut-off frequency for
spectra where {\nolinebreak{$\alpha\ge1$}} (see e.g.
\citet{barnes1971}). On the other hand, for
{\nolinebreak{$0\le\alpha<1$}} the power spectra must have an upper
cut-off frequency because of the same arguments. Thus, white noise in
this sense has to be ``band-limited'' which is automatically the case
in any real experiment due to inevitable time constants for example.
The special case of ``flicker noise'' ($\alpha=1$) requires both, a
lower and an upper cut-off frequency, in order to be realistic.
Consequently, the formulas (9) are valid within limits for $\tau$,
which are also defined by the appropriate cut-off frequencies.
Important for the following treatment is that for $1<\alpha\le3$
Eq.(9) is valid also for $\tau\rightarrow0$. The range $0<\alpha\le1$
we do not consider any further, since these noise power spectra don't
seem to be observable under normal circumstances, at least with
standard radio-astronomical equipment.

In this approximation we have now for the Allan variance according to
Eq.(6), (7), and (9):

\begin{equation}
\begin{array}{rcll}
  \sigma_{\rm A}^2(T) & = & 
    g_\alpha \frac{4 (2^{\alpha-1}-1)}{\alpha(\alpha+1)} 
    T^{\alpha-1} & 1<\alpha\le3 \\
                & = &
    g_0/T    & \alpha=0 \,\,{\rm{(white \,\,noise)}}  \nonumber \\
\end{array}
\end{equation}

 For $\alpha>1$ Eq.(10) is valid for integration times $T$ smaller than
 the characteristic correlation time of the drift noise and larger
 than is determined by the highest frequency components of the
 noise. These two assumptions apply in all cases considered here.

If we assume a simple power law for the drift contribution with a well
defined $\alpha$, and if we consider the additional presence of
radiometric noise, or ``white noise'', we expect the Allan variance to
have the following structure as a function of integration time:

\[
  \sigma_{\rm A}^2(T) = a/T + b T^\beta \,\,\,\,\,(\beta=\alpha-1).
\]

It is general experience with radio-astronomical as well as ordinary
laboratory equipment that the slope of the drift contribution is found
somewhere between $\beta=1$ and $\beta=2$, which corresponds to
$1/f^2$- and $1/f^3$-noise respectively. Good examples of such
correlation functions are the spontaneous decay of excited molecular
states with a simple exponential correlation function, or emission
from a thermal source with a Gaussian correlation function. When
expanded in lowest order approximation, they result in terms with
$\beta=1$ and 2 respectively. Chaotic processes will typically lead to
power-laws somewhere in between. We have never found an indication of
the presence of $1/f$-noise in any of our instruments which would
contribute with a horizontal slope in the Allan variance plot.

  Within the white noise part of the Allan plot, i.e. the regime with
  the slope of ``$-1$'', the radiometer equation must be valid:

\begin{equation}
  \sigma_{\rm A}^2(T) = \frac{\langle s(t)\rangle}{B_{\rm Fl} T}
\end{equation}

$B_{\rm Fl}$ is the ``fluctuation bandwidth'' of the spectrometer of the 
frequency channel of the spectrometer, which is defined as:

\begin{equation}
  B_{\rm Fl} = \frac{[\int_0^\infty P(f) df]^2}
                {\int_0^\infty P^2(f) df}
\end{equation}
(see e.g. \citet{kraus1980} and references therein). $P(f)$ is the power
response function of the frequency channel to a monochromatic input at
frequency $f$. $B_{\rm Fl}$ is always larger than the resolution-bandwidth
$\delta_{\rm Res}$ of the channel, so that the radiometric noise should be
somewhat smaller than often is expected. Typically $B_{\rm Fl}$ is more
than 50\% larger than $\delta_{\rm Res}$.

In most practical cases it is very useful to refer to the particular
integration time in the Allan variance plot where the minimum occurs.
This minimum describes the turn-over point where the radiometric noise
with a slope of $-1$ in the logarithmic plot becomes dominated by the
additional and undesired drift noise (see Fig.\,1). Above the minimum
time the rms of the measurements becomes much larger than is
anticipated by the radiometer equation alone. Intuitively, the minimum
time might appear as an upper limit for the integration on individual
positions during radio-astronomical observations, but the Allan
variance plot offers a lot more detailed advice when planning the most
efficient observing strategy under the given circumstances. Since any
additional noise above the radiometric level is very unfavorable, one
has to find the optimum integration time, where the loss due to
inevitable dead time during slew of the telescope etc. is as little as
possible, and where the impact of drift contributions is nearly
negligible at the same time. To find this best compromise is the goal
of the following chapters.

By use of the minimum time $T_{\rm A}$ of the variance we can now rewrite the
above equation with:

\begin{equation}
  \frac{\sigma_{\rm A}^2(T)}{\langle s(t)\rangle^2} 
   = \frac{1}{B_{\rm Fl} T_{\rm A}} 
     (1/t + t^\beta/\beta) \,\,{\rm{with}} \,\,
     t=T/T_{\rm A}
\end{equation}

In a mathematical sense the minimum time appears rather naturally as
the decisive parameter for the description of the plot. It is obvious
that at the minimum the variance is already significantly larger than
the radiometric value, for $\beta = 1$ it is doubled for example.

The slope of the drift part in the Allan variance plot is, as is seen
in Fig.\,1, also one of the important parameters for the
characterization of the instrument. Therefore, we can conclude that
the minimum time, the fluctuation bandwidth, and the slope at large
integration time are the three parameters which fully characterize the
instrument in a statistical sense. All three parameters are directly
accessible from the Allan variance plot once there are sufficient data
collected for a reliable evaluation. It is interesting to note that
generally the outcome of an Allan variance test looks nearly identical
to previous ones as long as the instrumentation used for the test is
not altered. This is particularly useful for checking the health of an
instrument from time to time. Certainly, there are other methods to
describe the noise performance of a radiometer like the plot of the
noise power spectrum or the correlation function or else, but it seems
rather natural to use the Allan variance plot, since it is directly
related to the normal observing procedure when observing an ``On''-
and an ``Off-position'' with a radio-telescope.

If the fluctuation bandwidth $B_{\rm Fl}$ is changed the minimum also shifts
due to the changing level of white noise, but, despite the change of
the leading factor, Eq.(13) is not altered due to the normalization
of the time with the Allan variance minimum time. How the radiometric
contribution is decreasing with increasing fluctuation bandwidth is
clear from the radiometer equation. However, the drift contribution
should not change, since it does not depend on the shape of the
filter-function of the actual spectrometer channel. The minimum
therefore shifts to smaller times with increasing $B_{\rm Fl}$ like
\begin{equation}
  T_{\rm A}' = T_{\rm A} (B_{\rm Fl} / B_{\rm Fl}')^{1/(\beta+1)}
\end{equation}
This formula should help when considering the stability of the
spectrometer output while co-adding adjacent pixels for example. (The
problem, how the fluctuation bandwidth changes when co-adding, is not
so easily solved. This is discussed in the appendix.)

Co-adding frequency pixels of a spectrometer output is standard
practice in radio-astronomy when dealing with very broad emission
lines e.g. from other galaxies. Thus it is not uncommon to finally
discuss spectra with an effective fluctuation bandwidth of the order
of 50 MHz by binning several spectrometer channels. A typical minimum
time of a complete radiometer system at an observatory is somewhere
around 30 seconds or so at a resolution of 1 MHz of the
spectrometer. According to Eq.(14) one would expect a shift of the
minimum time to values somewhere between 4 and 8 seconds for the
bins. A much larger bandwidth one has to deal with, when measuring
continuum signals with large bandwidth bolometers. A typical effective
bandwidth may be of the order of some 50 GHz. In this case the minimum
of the Allan variance moves to values between 0.1 and 0.8 seconds,
when assuming the origin of the white noise is still just radiometric
while the drift noise remains as before. It is clear that the
integration time used for sampling on each position may be a few
seconds in the first case, but has to be less than 100 msec in the
second. 

\section{Using the information contained in the Allan variance plot}

As was mentioned above, the Allan variance plot provides information
about what to expect in case there are no gaps in time between the
corresponding measurements ``signal'' (On) and ``reference'' (Off).
This is very close to the standard situation during observing, but now
the presence of dead time has to be included into the discussion. When
investigating the simple description of the Allan variance as a
function of integration time from above it seems plausible that the
plot should also provide all information about the impact of drift
noise, if there is dead time between the two measurements. How to do
this is fairly straightforward, and, in order to keep things short, we
present the mathematical treatment only briefly.

\subsection{Position-Switch observations}

Position-Switch measurements with one signal integration (On) per
reference measurement (Off) are very common for the observation of
single positions in an extended source for example. In other cases
Beam-Switch with a wobbling secondary mirror or Frequency-Switch
measurements are applied, since these methods seem to be more
promising for the resulting signal to noise ratio. In terms of a more
mathematical treatment, all these methods are identical, only the
typical time scale is different. In practice some dead time needs to
be included in the observing procedure, but both, On- and
Off-integration, are assumed to be of equal length.\footnote{The
assumption of equal length is only valid for identical noise levels of
both measurements $x_{s}$ and $x_{r}$. If the emission from the two
positions is very different and not small in comparison to the
receiver noise temperature, an equal length of the two integrations is
no longer a proper choice. This would apply when studying emission
from the sun for example, but in radio-astronomy, it would be an
exceptional situation.}  Following Eq.(1) we have for the signal- and
the reference-measurement:
\begin{eqnarray}
  x_{s}(T,t) & = & 1/T \int_{t-T}^t dt' s(t'), \nonumber\\
  x_{r}(T,t) & = & 1/T \int_{t+T_d}^{t+T_d+T} dt' s(t') \nonumber
\end{eqnarray}
when including the delay time $T_d$ between the end of the
On-integration and the begin of the Off-integration. For the error
estimate of difference of these two measurements we get now with the
help of Eq.(5), (6), and with $\sigma_1^2(T,T_d) = \Gamma_T(0) -
\Gamma_T(T_d+T)$ (similar as in Eq.(7)):

\begin{equation}
  \frac{\sigma_1^2(T,T_d)}{\langle s(t) \rangle^2}
  = \frac{2}{B_{\rm Fl} T_{\rm A}} 
    (1/t + 1/\beta f(t,d)), 
\end{equation}
$t=T/T_{\rm A}$, $d=T_d/T_{\rm A}$ with \newline
$f(t,d) = t+3/2 d$, $\beta=1$, and \newline
$f(t,d) = [t+d]^2$, $\beta=2$.

\vspace*{0.2cm}

It is possible to derive suitable expressions for arbitrary values of
$\beta$, but in the following treatment we concentrate on the two
extreme cases $\beta=1$ and 2 only. $\sigma_1^2(T,T_d)$ describes now
the noise found with one single pair On and Off. It is most efficient
to move the telescope only every second time so that the observing
sequence is On--Off/Off--On/On--Off ...  instead of
On--Off/On--Off/On--Off....  (This is also true for Beam-Switch
measurements!) In this case we have for the duration of each complete
cycle with one On- and one Off-integration:

\[
  T_{c} = 2 T + T_{d}
\]

Usually, the measurement is repeated several times and the result is
co-added to improve the signal to noise ratio. Then we have $K$ such
pairs, which are measured within a total observing time $T_{\rm Obs}$. We get
therefore for a given observing time $T_{\rm Obs}$: 

\[
  T_{\rm Obs} = K T_{c}.
\]

Since the variance should develop like $1/K$, we have finally for the
variance of the complete observation on one On-position.\footnote{At
long total observing time the reduction of the variance like $1/K$ can
be proven for any realistic noise power spectrum when using the fact
that the noise correlation function must stay finite for
$\tau\rightarrow0$ (see also above).}
\begin{eqnarray}
  \lefteqn{
  \frac{\sigma_{\rm K}^2(T,T_d)}{\langle s(t) \rangle^2}
  = 1/K \frac{\sigma_1^2(T,T_d)}{\langle s(t) \rangle^2} 
  }\\
  & & = \frac{4}{B_{\rm Fl} T_{\rm Obs}}  
    (1/t + f(t,d)/\beta) 
    (t+d/2) \nonumber
\end{eqnarray}

Any realistic drift scenario can be described by this formula, and the
result must be located within the range of the two limiting values of
$\beta$. For a useful calculation it is now mandatory that the
information about the minimum time $T_{\rm A}$ is known from an Allan
variance measurement.

Fig.2 shows the shape of Eq.(16) as a function of the relative
integration time $t$ for a few values of $d$. For each $d > 0$ the
function has exactly one fairly broad minimum, and it is plausible
that only in this minimum the observation can be done with maximum
efficiency. Any other $t$ leads to a higher noise level, i.e. to lower
efficiency within a given observing time. This can be explained by the
facts that with very short integration a lot of time is wasted while
moving the telescope, and that at very long integration time the drift
noise starts to deteriorate the signal to noise ratio on the other
hand. In Fig.3 the optimum integration time at the minimum of the
variance is shown for both cases $\beta=1$ and 2 as a function of the
relative dead time $d$. The preferred relative integration time $t$ is
always significantly smaller than unity, which leads to the important
conclusion that the integration time should always be considerably
smaller than the Allan variance minimum time. With a realistic drift
noise contribution ($1\le\beta\le2$) the optimum integration time will
be located somewhere between the two solid lines in the plot. For the
figure, also those limits for the integration time have been computed,
where the rms-noise is increased by less than 1\% as compared to the
optimum. The dotted curves indicate these limits for both $\beta$, and
it is appears that these regions overlap largely. The hatched area in
the plot indicates where this overlap-region is found. It means that
for any realistic scenario it is always possible to find an
integration time with almost perfect noise performance independent on
the actual drift characteristics of the system. Consequently, the
precise knowledge of the drift slope $\beta$ is not really essential
for the optimization procedure.

\begin{figure} [h!]
\resizebox{\hsize}{!}{\includegraphics{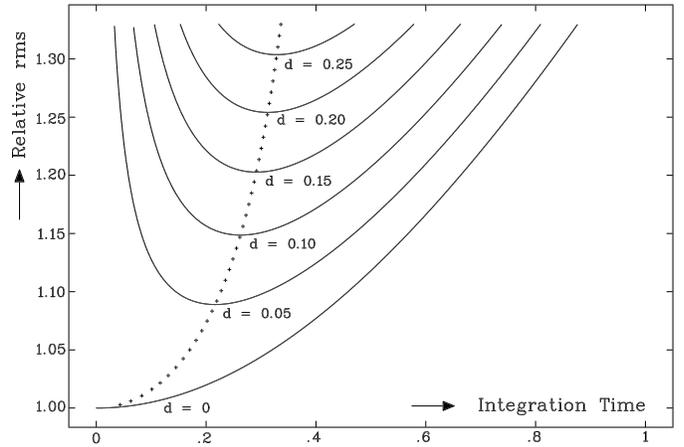}}
%\resizebox{\hsize}{!}{\includegraphics{2test.eps}}
% \resizebox{6cm}{!}{\includegraphics{ci_map_spec_x.eps}}
\caption{ \label{FIG_SPECTRA2}
  The development of the rms of Position-Switch measurements as a
  function of integration time for a drift slope of $\beta=1$ in the
  Allan variance plot (see Eq.(16)). The curves are calculated for
  several delay times between On- and Off-position ($d$ = 0,...,0.25).
  The dotted curve connects all minima of the curves and represents
  the optimum integration time for all delays. The values of the delay
  time $d$ as well as of the integration time are given in units of
  the Allan variance minimum time.  }
\end{figure}

\begin{figure} [h!]
\resizebox{\hsize}{!}{\includegraphics{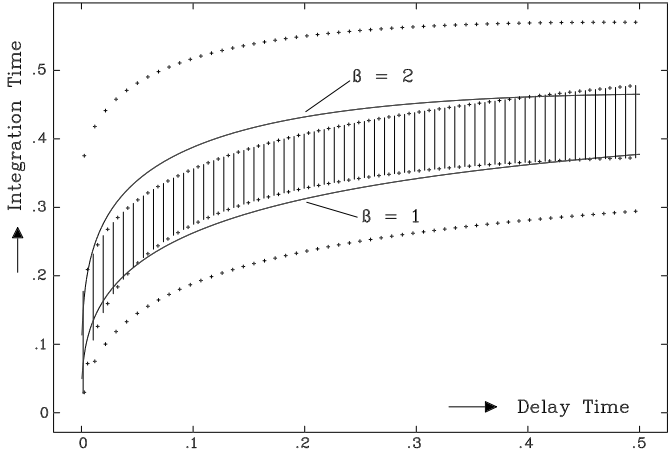}}
% \resizebox{6cm}{!}{\includegraphics{ci_map_spec_x.eps}}
\caption{ \label{FIG_SPECTRA3}
   Optimum integration time as a function of On-Off delay for the two
   extreme drift contributions with $\beta=1$ and $\beta=2$ as found
   from Eq.(16).  The dotted curves represent the intervals where the
   rms is increased by 1\% maximum for both values of $\beta$. The
   hatched area defines the regime where the rms increase is less than
   1\% independent on the actual value of $\beta$. In this area the
   preferred choice of the integration time is found. The values of
   the delay time $d$ as well as of the integration time are given in
   units of the Allan variance minimum time. }
\end{figure}

 As was mentioned before, with a standard low resolution spectrometer
 one typically finds an Allan variance minimum of a complete
 radiometer system in the range of 30 seconds or so. Chopped
 measurements, using a wobbling secondary telescope mirror for
 example, are considered as the ideal method for point-like sources to
 reduce the impact of drift noise on the appearance of the baselines
 of the spectra. If the chop delay, i.e. the time to move the
 subreflector between the two positions, needs 100 msec for example,
 the optimum integration time per position is found near 4 seconds
 following Eq.(16). The situation seems to be different for the case
 $d = 0$, as it would apply for Frequency-Switch measurements for
 example, since the switch between the two nearby frequencies takes
 negligible time. But, as is visible in Fig.2, the increase in rms
 noise is fairly marginal ($\le1$\%) even for integration times $T$ up
 to 14\% of $T_{\rm A}$.  This means, in all practical cases it is of no use
 to switch at high speed, on the contrary, the efficiency of the
 observation might become affected, if dead time is involved. Even for
 spectra at moderately reduced frequency resolution the required
 integration time does not drop significantly below 1 second. It is
 therefore important to note, that a higher chop frequency is only
 required for continuum measurements with very large bandwidth. 

The ideal, theoretical limit for the observing efficiency is reached,
when effectively all integration time is spent on the On-position and
if there would be no dead time involved. In this case we have:

\[
  \frac{\sigma_{\rm th}^2(T)}{\langle s(t)\rangle^2} = 
  \frac{1}{B_{\rm Fl} T_{\rm Obs}}
\]

The best possible efficiency relative to this theoretical performance
is therefore:
\begin{eqnarray}
  \eta & = & [\sigma_{\rm th}^2(T)/\sigma_K^2(T,T_d)]^{1/2} \\
       & = & 1/2 [(1/t_0+1/\beta f(t_0,d))
                     (t_0+d/2)]^{-1/2} \nonumber
\end{eqnarray}
with $t_0=T_0/T_{\rm A}$ the optimum integration time for the given
delay. This observing efficiency $\eta$ is always smaller than 50\%,
since at least half of the time is ``wasted'' for the integration of
the Off-signal. Clearly, the longer the dead time the less efficient
the observation. Since the impact of the dead time is determined by
its relative length when comparing with the Allan variance minimum
time, a larger $T_{\rm A}$ helps as well. (A plot of Eq.(17) can be found
in Fig.5.)  It should be kept in mind that the efficiency calculated
here is the best possible for a given $d$. If other integration times
are chosen, the efficiency will definitely become worse! One should
also be aware of the fact that the total observing time has to be
increased by a factor proportional to the square of the inverse
efficiency to compensate for the reduced efficiency, which might
become a high price to pay for a non-appropriate observing strategy.

\subsection{Mapping}

Another and possibly more interesting case is the situation when
measuring maps either by Raster-Mapping or On-The-Fly. In both cases
there are $N$ different On-positions per Off-position in one cycle,
the only difference is that for Raster-Mapping there is some dead time
between the different On-positions which does not appear during OTF
observations. It is found in literature that the Off-integration time
should be $\sqrt{N}$ times longer than the On-integration time
\citep{ball1976}. This advice leaves the question open how long the
On-integration should last. For the following treatment of this
question we assume that we have an On-integration time $T_{s}$, an
Off-integration time $T_{r}$, a dead time $T_{ds}$ between each of the
On-measurements, another dead time $T_{dr}$ to move from the last On-
to the Off-position, and a different dead time $T_{dc}$ to move the
telescope back to the first On-position to begin with the next cycle
again. It is plausible that $T_{dc}$ will not be identical with
$T_{dr}$, since the first and last On-position are not the same, and
the time to move between the positions (with different velocity
requirements in OTF-mode as well) is definitely different.

The delay between one of the On-positions and the Off-position is also
dependent on the number of Ons in between. If we consider the worst
case situation, we have to investigate the On-Off pairs with maximum
delay involved, which is the first On-position when putting the Off at
the end of the cycle. The delay $T_d$ is then:

\begin{equation}
  T_d = (N-1) (T_{s}+T_{ds}) + T_{dr} \,\, {\rm{or}} \,\,
  d = (N-1) (s+d_{s}) + d_{r}
\end{equation}

Here and for the following we use $d=T_d/T_{\rm A}$, $d_{s}=T_{ds}/T_{\rm A}$, 
$d_{r}=T_{dr}/T_{\rm A}$, $d_{c}=T_{dc}/T_{\rm A}$, $s=T_{s}/T_{\rm A}$ and $r=T_{r}/T_{\rm A}$.

We also have to take into account now that the integration time for On
is different than for Off. Hence we write:
\begin{eqnarray}
  x_{s}(T_{s},t) & = & 1/T_{s} \int_{t-T_{s}}^t s(t') dt', \\
  x_r(T_{r},t) & = & 1/T_{r} \int_{t+T_d}^{t+T_d+T_{r}} s(t') dt' 
  \nonumber
\end{eqnarray}

Similar as before we find after some straight-forward derivation using
Eq.(5),(6),(9), and (19):

\[
  \sigma_1^2(s,r) = \Gamma_{T_{r}}(0) + \Gamma_{T_{s}}(0)
  - [w_+\Gamma_{T_+}(T_{m}) - w_-\Gamma_{T_-}(T_{m})]
\]
with $T_\pm = (T_{r}\pm T_{s})/2$, $w_\pm=2 T_\pm^2/[T_{r} T_{s}]$,
and $T_{m}=T_+ +T_d$.  When integrating one finds now:

\begin{equation}
  \frac{\sigma_1^2(s,r)}{\langle s(t)\rangle^2} 
  = 
  \frac{1}{B_{\rm Fl} T_{\rm A}} 
  (1/s + 1/r + 2g(s,r,d)/\beta)
\end{equation}

and for the two limiting cases of $\beta$ one gets:\newline
$g(s,r,d)=(s+r)/2+3/2 d$, $\beta=1$, and \newline
$g(s,r,d)=[(s+r)/2+d]^2$, $\beta=2$.

\vspace*{0.2cm}

The function $g(s,r,d)$ is identical with $f(t,d)$ for $s=r=t$ (see
Eq.(15)). The variance found here is valid for one pair of a
particular On- and the corresponding Off-measurement.

We have to identify now, how the noise is developing, if one wants to
observe a full map within a given total observing time $T_{\rm Obs}$. One
observing cycle consists of $N$ identical On-integrations ($T_{s}$), one
Off-integration ($T_{r}$), and the various dead times in between. Thus
we have for the complete cycle time $T_{c}$:

\begin{equation}
  T_{c} = N T_{s} + T_{r} + (N-1) T_{ds} + T_{dr} + T_{dc}
\end{equation}

We assume that we want to measure a map consisting of $L$ different
On-positions. This needs $L/N$ cycles for observing each position
once. Each of the On-positions may be measured $K$ times within the
total observing time $T_{\rm Obs}$ in order to improve the noise
level. Thus we have:

\begin{equation}
  T_{\rm Obs} = K T_{c} \times L/N.
\end{equation}
with $K\ge1$. The choice of $K$ may be dependent on $N, L$ and the
available total observing time $T_{\rm Obs}$, and it has to be chosen
according to the individual needs of the observing program. In many
cases $K$ will be equal to 1. When using Eq.(20), (21), and (22), we
get now finally:
\begin{eqnarray}
  \lefteqn{
  \frac{\sigma_K^2(s,r,N)}{\langle s(t) \rangle^2} =
  1/K \frac{\sigma_1^2(s,r,N)}{\langle s(t) \rangle^2}
  }\\
  & & = L \frac{1}{B_{\rm Fl} T_{\rm Obs}} (1/s+1/r+2/\beta g(s,r,d)) \nonumber\\
  & & \hspace*{1.5cm} \times (s+d_{s}+\frac{r+d_{r}+d_{c}-d_{s}}{N}) \nonumber
\end{eqnarray}

We have found now the variance as a function of three variables $s$,
$r$, and $N$ with the relative delays $d_{s}$, $d_{r}$, and $d_{c}$ as
parameters. Note that the On-Off delay $T_{dr}$ has different impact
on the statistics than the return delay $T_{dc}$, since the latter
does not affect the drift contribution $g(s,r,d)$ (see Eq.(18), (20),
(21), and (23)).

The minimum of $\sigma_K^2(s,r,N)$ can be found, where all derivatives
with respect to $s$, $r$, and $N$ become zero. This is the set of
variables where the observing efficiency becomes the best possible
under the given circumstances. (It is simple to prove that there is
exactly one minimum as long as $s$, $r$ and $N$ are larger than zero.)
Any other set of variables will result in a degradation of the
observing efficiency. But, as was mentioned before, the use of the
relation $r=s\sqrt{N}$ leads to results very close to this
optimum.\footnote{Using Eq.(23) it is easy to verify this relation
  when assuming that there is no drift contribution involved. But, if
  there is drift noise, it is also clear from Eq.(23) that the
  relation is no longer valid. However, a comparison of the results of
  a calculation with and without the relation between On- and Off-time
  shows that the minimum rms-values differ only by amounts of the
  order of 0.1\% or less. Therefore the introduction of the simple
  relation between $s$ and $r$ remains justified.}  Therefore, for all
practical purposes it is sufficient to apply only a two-dimensional
optimization for the two variables $s$ and $N$:
\begin{eqnarray}
  \partial\sigma_K^2(s,r,N)/\partial s|_{r=s\sqrt{N}} & = & 0 \,\,{\rm{and}} \\
  \partial\sigma_K^2(s,r,N)/\partial N|_{r=s\sqrt{N}} & = & 0 \nonumber
\end{eqnarray}

It is trivial to show that the optimum number of Ons becomes infinite
in case of OTF measurements ($d_s=0$). Therefore it seems to be
advisable to use fairly large $N$ in order to be as close as possible
to the optimum case of $N\rightarrow\infty$. On the other hand, the
optimum integration time $t_s$ becomes extremely small in this case (see
below), which finds it's limitation because of hardware constraints
for example. Surprisingly, for Raster-Mapping with $d_s\neq0$ there is
always a finite $N$ required for an optimized observation. This optimum
$N$ is dependent on $d_s$, $d_r$, and $d_c$.

Usually, it is rather difficult to make observations with an arbitrary
number of Ons per Off at a given geometry of a particular map. It is
therefore much more interesting to derive conclusive estimates for an
optimized observation under the assumption of a predefined and fixed $N$
for both, Raster-Mapping and OTF observations. In this case one has to
find the minimum with:
\begin{equation}
  \partial\sigma_K^2(s,r,N)/\partial s|_{N\,{\rm{fixed}}, r=s\sqrt{N}} = 0
\end{equation}

In any case one has to investigate what impact the chosen $N$ has on
the total efficiency using Eq.(23) and (24) in order to verify that
the used $N$ is not too far away from optimum.

In order to provide some idea about the best choice of the
On-observing time $s$, the optimum integration time in OTF mode is
shown in Fig.4 as a function of the On-Off delay $d_r$. The delay for
the return to the begin of the cycle is taken into account by a $d_c$
20\% longer than $d_r$. The two solid curves are derived from Eq.(23)
and (20) for the two limiting cases $\beta=1$ and $\beta=2$.  The
hatched area in the plot represents the region where the increase of
the rms stays below 1\% as compared to the optimum for both values of
$\beta$. This means that for all assumed drift slopes one is always
safe when choosing an On-integration time within this region.  Such
optimized integration time can be described by the purely empirical
formula:

\begin{equation}
  s \approx 0.53 d^{0.23}/N^{0.69}, r=s\sqrt{N}
\end{equation}
with $d=(N-1)d_s+d_r+d_c$.

$d$ is the sum of all delays in one cycle. The formula is also valid
for Raster-Mapping and Position-Switch measurements, and it may be
used for values of $d_r$ and $d_c$ between 0 and 1, for $d_s\le0.1$,
and $N\ge1$.

%Fig.4
\begin{figure} [h!]
\resizebox{\hsize}{!}{\includegraphics{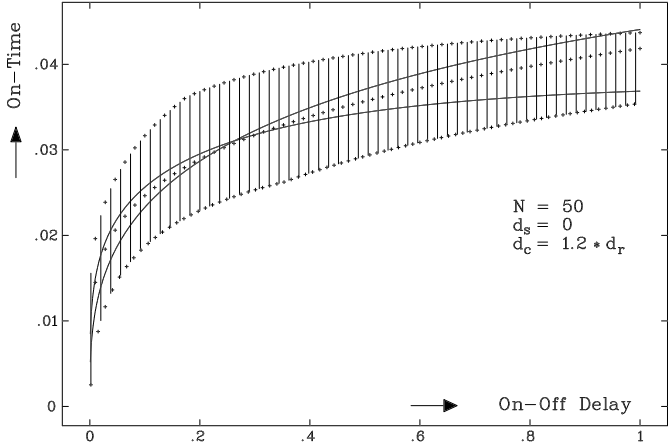}}
% \resizebox{6cm}{!}{\includegraphics{ci_map_spec_x.eps}}
\caption{ \label{FIG_SPECTRA4}
  Optimum On-integration time for OTF measurements with 50 Ons per
  Off. The hatched area represents the regime where the rms increase
  stays below 1\% for any $\beta$ between 1 and 2. The dotted curve in
  the middle represents the suggested On-integration time using
  Eq.(26). As is clearly visible, the optimum integration time is
  typically of the order of a few seconds when assuming an Allan
  variance minimum time near or above 100 seconds.  }
\end{figure}

Finally, also the overall observing efficiency can be found for the
measurement of extended maps. The theoretically best possible value of
the variance is given by:

\[
  \frac{\sigma_{\rm th}^2(s,r,N)}{\langle s(t) \rangle^2} = 
  L \frac{1}{B_{\rm Fl} T_{\rm Obs}},
\]
where no dead time is present and virtually all observing time is spent on
the On-positions. In this case we have now for the relative efficiency:
\begin{eqnarray}
  \lefteqn{
  \eta = [\sigma_{\rm th}^2(s,r,N)/\sigma_K^2(s,r,N)]^{1/2} 
  }\\
  & & = \Bigl[ 
  (1/s+1/r+2/\beta g(s,r,d)) \nonumber \\
  & & \hspace*{1.5cm} \times (s+d_s+\frac{r+d_r+d_c-d_s}{N})
        \Bigr]^{-1/2} \nonumber
\end{eqnarray}

Fig.5 depicts the optimum efficiency according to Eq.(27) and (25)
for three different $N$ ($N$ = 1, 10, and 100). The curves for $N=1$
(dotted lines) are the Position-Switch efficiencies at the same time
(see Eq.(17)). Clearly the OTF efficiency is much better than the
Position-Switch efficiency. At zero delay it reaches a maximum value
of $(1+1/\sqrt{N})^{-1}$, and it decreases monotonically with
increasing $d_r$. Again, the efficiency shown in the plot is the
maximum one can achieve under the given circumstances. When comparing
$\eta(N=10)$ with $\eta(N=100)$, it is clear that $N=100$ is the
preferable choice. This example demonstrates that it is advisable to
determine whether the number of desired $N$ is a reasonable choice or
should be reconsidered when planning the best strategy for the
observation.

%Fig.5
\begin{figure} [h!]
\resizebox{\hsize}{!}{\includegraphics{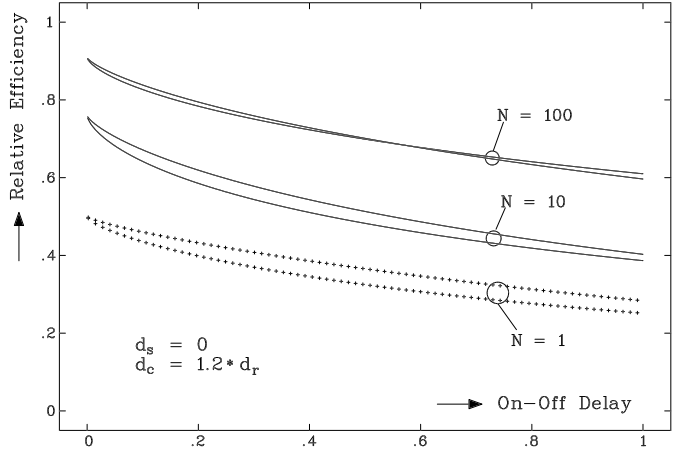}}
% \resizebox{6cm}{!}{\includegraphics{ci_map_spec_x.eps}}
\caption{ \label{FIG_SPECTRA5}
  Relative optimum efficiencies of OTF measurements for $N$=1, 10, and
  100 On-positions per Off (see Eq.(27)). For each $N$ both curves for
  $\beta$ = 1 and = 2 are plotted. It is obvious that larger $N$ lead to
  higher efficiency. The dotted curves for $N=1$ represent the
  Position-Switch situation with an On-Off delay every second time
  only. This is taken into account by setting $d_c = d_s = 0$ in Eq.(23)
  and (27) while $N = 1$.  }
\end{figure}

How the efficiency develops with $N$ is visible in Fig.6 for some fixed
On-Off delays. Obviously, the gain in efficiency with increasing $N$
above $N = 50$ is rather marginal. Therefore it is questionable whether
a significant improvement in observing efficiency is achievable when
going from $N = 50$ to $N = 100$ for example. Any reduction of the On-Off
delay time would be a much more effective measure. On the other hand,
the plot shows also, how valuable an increase in $N$ can be in case one
is considering $N = 10$ or less. 

%Fig.6
\begin{figure} [h!]
\resizebox{\hsize}{!}{\includegraphics{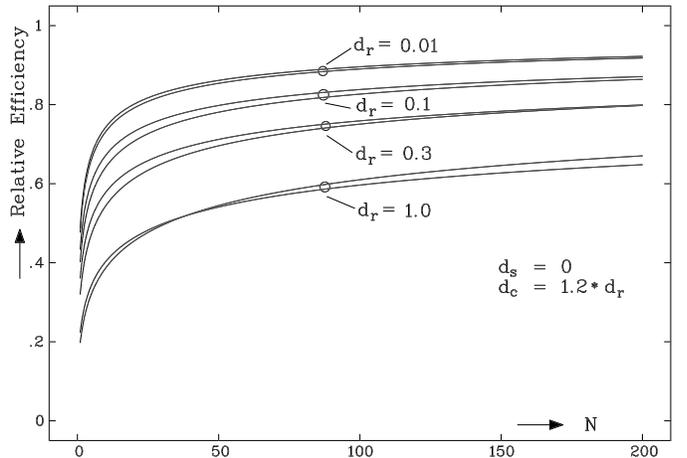}}
% \resizebox{6cm}{!}{\includegraphics{ci_map_spec_x.eps}}
\caption{ \label{FIG_SPECTRA6}
  Relative OTF efficiency as a function of the number of Ons per Off for
various relative On-Off delays according to Eq. (27), (25), and
(23). For each $d_r$ both curves for $\beta = 1$ and $= 2$ are plotted.  
  }
\end{figure}

One of the remaining questions is, how long one cycle $T_c$ will last,
once the optimum On- and Off-integration time has been found. Using
Eq.(21) it is now simple to calculate $T_c$ as a function of the
On-Off delay time $d_r$. In Fig.7 the cycle time is plotted for three
cases with $N$ = 1, 10, and 100. At first sight it appears surprising
that the time for a full cycle increases to values several times
longer than the Allan variance minimum time in case there is
substantial delay $d_r$. But again, the length of one cycle depends
strongly on the number of Ons per Off. Since the On-integration time
is rather small at large $N$, the larger radiometric noise of the
On-measurement dominates the noise budget so that a longer delay with
an increased contribution of drift noise becomes acceptable. For a
given and fixed $N$ the increase of the cycle time with increasing delay
is the consequence of the fact that at larger integration time the
loss due to drift noise is less costly than the loss due to the On-Off
delay. This effect is also clearly visible in Fig.2 for the case of
Position-Switch measurements.

%Fig.7
\begin{figure} [h!]
\resizebox{\hsize}{!}{\includegraphics{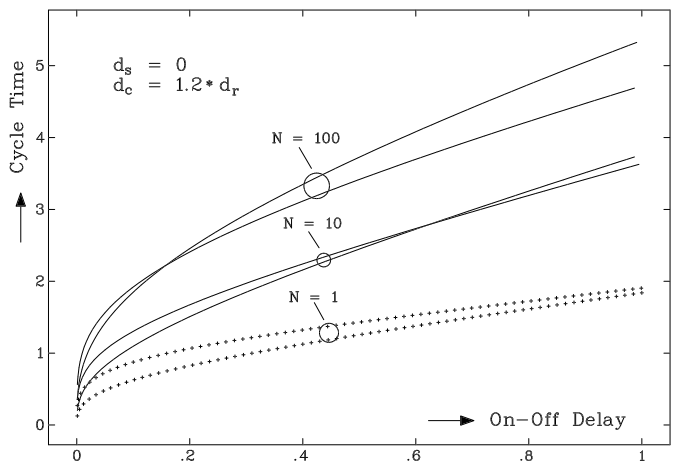}}
% \resizebox{6cm}{!}{\includegraphics{ci_map_spec_x.eps}}
\caption{ \label{FIG_SPECTRA7}
Cycle time for OTF measurements as a function of On-Off delay. 
%ckr: added (6.1.2000)
The cycle time comprises $N$ On-integrations, one Off-integration, and
the dead times in between.
%ckr.
The three cases ($N$=1, 10, and 100) are calculated from Eq.(21),
(23), and (25). Similar to Fig.5, the Position-Switch situation is
also indicated by the dotted lines. Note that the increase of cycle
time is partly due to the time spent during slew from On to Off and
back. }
\end{figure}

\section{Conclusion}

The discussion above provides some clear guidelines for an optimized
observing program. The first step has to be a reliable measurement of
the system Allan variance. The word "system" includes all components
of the observatory which may possibly contribute to the noise
including the atmospheric fluctuations for example. When knowing the
applicable dead times, a simple calculation of the optimum integration
time can be made by using the "rule of the thumb" as given by
Eq.(26). As was pointed out before, Position-Switch or Chop
measurements should be done in a most economical way by moving the
telescope or the chopper only every second time. OTF or Raster-Mapping
measurements need a clear understanding of the impact of the number of
On-positions chosen for each Off-integration.  Also here it might be
of some value to reverse the sequence of the integrations on the
various positions every second time in order to reduce some of the
loss in time due to the slew of the telescope between the On- and the
Off-positions.  It should be noted that the measurement of large maps
can be handled in different ways. If one wants to achieve a certain
signal to noise, it might be advisable to use larger $N$ with smaller
$T_s$ and to repeat the map several times, as it is considered by the
parameter K in Eq.(22). In any case, the suggested On- and
Off-integration time should not be drastically altered, although the
plot in Fig.4 indicates that there is quite some margin available.

In general it is surprising how closely together the curves for the
different $\beta$ in Figs.5, 6, and 7 are found, which is a clear
validation for the assumption that it is sufficient to consider only
the extreme cases for the drift contributions. Therefore, there is no
need to go too deeply into the analysis of the drift part in the
noise. It is also one of the better news from the treatment here that
some freedom to plan the observation is still preserved. This might be
particularly important when considering the constraints set by the
observatory hardware. It is probably not advisable to operate with too
short integration intervals, since the data flood might become
overwhelming, and the storage capacity of the computers could easily
be exceeded. Therefore, the conclusion found before that there are no
real requirements for high speed observing most of the time is very
important.

The discussion above is most useful for observations with space-born
observatories like SWAS \citep{melnick2000}, ODIN
\citep{hjalmarson1993} or FIRST \citep{graauw1998}. \footnote{FIRST
  was recently renamed to ``Herschel Space Observatory''.} Since
  usually a satellite cannot be oriented in space very rapidly, the
  impact of dead time becomes vital. The SWAS satellite is not capable
  to control the pointing very accurately during slew across an
  extended source, so that the OTF mode is not applicable. Instead,
  Raster-Mapping is a generally used procedure. On the other hand,
  since SWAS is a very small satellite, it can be pointed from one
  position to a second in 3 degrees distance within less than 15
  seconds. A 3-degree nod is often required during observations in the
  Milky Way, since the emission of molecules like CO is fairly
  extended. Nevertheless, the loss in observing efficiency looks
  acceptable, when considering an Allan variance minimum time of the
  SWAS receiver/backend system of about 150 seconds as found in orbit.
  On the Herschel space observatory, the situation will be changed
  drastically. We can assume that the pointing of the telescope during
  slew is well defined so that OTF measurements should be applicable.
  But, due to the fact that Herschel is going to be a very heavy
  satellite, the movement by three degrees will last nearly as long as
  the expected Allan variance minimum time will amount to. In
  consequence, the value of the dead times $d_r$ and $d_c$ will be
  close to unity when assuming a similar system stability like that of
  SWAS.  This prohibits Position-Switch measurements with the
  instrument, because the efficiency would drop to values below 30\%,
  which would certainly be rather disappointing because of the
  consequences for the extremely precious and limited observing time.
  Therefore, a very careful analysis for determining the best possible
  observing strategy is extremely important for such a program.

Rather different circumstances exist at ground-based
observatories. Typical dead time for a slew of 3 degrees is of the
order of a few seconds only, therefore the impact of dead time does
not appear as devastating as with space-based observatories. A
detailed planning of an observing strategy does not seem to be so
easily implemented, particularly, if other parameters like varying
hardware constraints or human limitations are playing a significant
role as well. Typically, the Allan variance minimum time of most
ground-based sub-millimeter observatories is rather small, partly due
to the impact of an unstable atmosphere. Therefore, the advantage of a
smaller dead time is partly eaten away by the reduced stability. But
still, as should be clear from the discussion before, the actual
situation has to be analyzed in detail for every individual case in
order to achieve as much scientific return from the observations as
possible. For this the usage of the analysis presented in this paper
could be very essential.

\appendix

\section{The development of noise when co-adding frequency pixels}

Co-adding a couple of pixels in a measured spectrum in order to
improve the signal to noise ratio is general practice when dealing
with noisy spectra, but, the consequences of this procedure are not
quite as trivial as one would like to believe. For the discussion we
start again with the definition of the normalized first order
correlation function as defined in Eq.(4): 
\[
  g_m = \langle dy_n dy_{n+m}\rangle/
  [\langle dy_n^2\rangle  
   \langle dy_{n+m}^2\rangle
  ]^{1/2}
\]
with $dy_n=y_n-\langle y \rangle$.

The data $y_n$ are here the pixel components of a fully calibrated
spectrum as measured with a multi-channel spectrometer. The index "m"
describes, by how many pixels the spectrum is shifted before the
multiplication of the pixel data is done.\footnote{In case of a finite
data set with $N$ data we can convert the definition into a more
practical definition using:

\[
  g_m = \frac{\frac{1}{N-m-1}
              \sum_{n=1}^{N-m}\delta y_n \delta y_{n+m}}
             {\Bigl( 
 \frac{1}{N-m-1} \sum_{n=1}^{N-m}\delta y_n^2 
 \frac{1}{N-m-1} \sum_{n=1}^{N-m}\delta y_{n+m}^2
             \Bigr)^{0.5}}
\]
with $\delta y_n = y_n -1/(N-m) \sum_{k=1}^{N-m} y_k$ and
$\delta y_{n+m} = y_{n+m} -1/(N-m) \sum_{k=1}^{N-m} y_{k+m}$. The
expectation values are estimated here by the means over a sufficiently
large number of data ($=N-m$). Important is to note that the value of
this auto-correlation function is ``1'' for $m=0$ by definition.
} 
The correlation function is symmetric,
since $g_{-m}=g_m$. We assume that all $y_n$ behave identically in a
purely statistical sense. Then, the values of $g_m$ depend only on the
``distance'' between the data given by the parameter ``m'', and the
expectation values as defined by the brackets become independent on
$n$. We have to determine now the expected statistics of the new
co-added data set $z_n$ with:

\[
  z_n = 1/K \sum_{k=1}^{K} y_{n+k}
\]
with $K$ the number of co-added pixels. With the usual definition of
the variance, $\sigma_K^2 = \langle z_n^2 \rangle-\langle z_n \rangle^2$,
we can now determine how the error of the new data develops:
\begin{eqnarray}
  & & \sigma_K^2 = \langle [1/K \sum y_{n+k}]^2\rangle - 
                   \langle 1/K \sum y_{n+k}\rangle^2 \nonumber \\
  & & = \langle [1/K \sum dy_{n+k}]^2\rangle \nonumber \\
  & & = 1/K^2 \sum_{p=1}^K \sum_{q=1}^K \langle dy_{n+p} dy_{n+q}\rangle
   \nonumber \\
  & & = \sigma_1^2/K^2 \sum_{p=1}^K \sum_{q=1}^K g_{p-q} 
      = \sigma_1^2/K [1+2\sum_{m=1}^{K-1}(1-m/K)g_m] \nonumber 
\end{eqnarray}

$\sigma_1^2$ is the variance of the statistical distribution of the
initial data $y_n$. From this and the radiometer equation we get now
finally:

\[
  \sigma_K^2 = \langle z \rangle^2/[B_K T] = \sigma_1^2/K_{\rm Box}
  = \langle y \rangle^2/[K_{\rm Box} B_1 T]
\]
with $K_{\rm Box}=K/(1+2(1-1/K)g_1+2(1-2/K)g_2+...)$. The new
fluctuation bandwidth $B_K$ is therefore $K_{\rm Box}$ times larger
than the fluctuation bandwidth $B_1$ of a single spectrometer pixel.
But, the effective number of pixels $K_{\rm Box}$ is significantly
smaller than the number of co-added pixels since the values of the
auto-correlation function are all positive under normal circumstances.
Note that the ratio of $K$ and $K_{\rm Box}$ is a function of $K$
itself so that one has to analyze the situation for the individual
case accordingly.

Only the first few values of $g_m$ ($m$ not larger than about 3)
should be non-zero for a decent spectrometer, since the overlap of the
power response functions between neighbored pixels should be small.
Therefore, in the limiting case of very large width of the bins ($K$
large), we get now:

\[
  K_{\rm Box} \approx K/(1+2g_1+2g_2+2g_3)
\]

Typical values for $K_{\rm Box}$ at large $K$ - for instance at
Nyquist sampling of the spectrum - are somewhere near $K/2$ depending
on the actual spacing and shape of the spectrometer channels, but they
may vary for different spectrometer types.

%%%%%%%%%%%%%%%%%%%%%%%%%%%%%%%%%%%%%%%%%%%%%%%%%%%%%%%%%%%%%%%%%%%%%%%%%%%

%\begin{acknowledgements}
%\end{acknowledgements} 

%\bibliographystyle{aabib99}    % aabib99.bst
%\bibliography{aamnem99,allan}  % aamnem99.bib, kramer.bib

\bibliographystyle{apj} % style file apj.bst, 3/2001
\bibliography{allan} % allan.bib

\end{document}